# Dzyaloshinskii–Moriya interaction and spin-orbit torque at the Ir/Co interface

Yuto Ishikuro[1], Masashi Kawaguchi[1], Naoaki Kato[1], Yong-Chang Lau[1,2], and Masamitsu Hayashi[1,2*]

[1]*Department of Physics, The University of Tokyo, Bunkyo, Tokyo 113-0033, Japan*

[2]*National Institute for Materials Science, Tsukuba 305-0047, Japan*

We studied the spin torque efficiency and the Dzyaloshinskii–Moriya interaction (DMI) of heterostructures that contain interface(s) of Ir and Co. The current-induced shifts of the anomalous Hall loops were used to determine the spin torque efficiency and DMI of [Pt/Co/X] multilayers (X=Ir, Cu) as well as Ir/Co and Pt/Ir/Co reference films. We find the effective spin Hall angle and the spin diffusion length of Ir to be ~0.01 and less than ~1 nm, respectively. The short spin diffusion length and the high conductivity make Ir an efficient spin sink layer. Such spin sink layer can be used to control the flow of spin current in heterostructures and to induce sufficient spin-orbit torque on the magnetic layer. The DMI of Ir and Co interface is found to be in the range of ~1.4 to ~2.2 mJ/m$^2$, similar in magnitude to that of the Pt and Co interface. The Ir/Co and Pt/Co interfaces possess the same sign of DMI, resulting in a reduced DMI for the [Pt/Co/Ir] multilayers compared to that of the [Pt/Co/Cu] multilayers. These results show the unique role the Ir layer plays in defining spin-orbit torque and chiral magnetism in thin film heterostructures.

*email: hayashi@phys.s.u-tokyo.ac.jp

## I. Introduction

The Dzyaloshinskii–Moriya interaction (DMI) emerges at interfaces between heavy-metal (HM) and ferromagnetic-metal (FM) layers and stabilizes chiral magnetic textures[1-3]. Chiral domain walls[4-7] and magnetic skyrmions[8-11] have been observed in systems with large DMI. In particular, the Pt/Co interface[4, 5, 8-11] is used as a platform to study static and dynamic properties of chiral magnetic textures owing to its large DMI.

The sign (i.e. chirality) and strength of interfacial DMI depend on the combination of materials. Although the exact microscopic origin[12-15] of the interfacial DMI remains to be identified, significant effort has been placed to develop heterostructures with large DMI. For example, the overall DMI of the system can be increased by sandwiching a FM layer with HM layers that induce opposite magnetic chirality. Typical examples of such structures are films that consist of Pt, Co and Ir. Experimental results[16, 17] and calculations[13, 14, 18] suggest that the magnetic chirality at the Ir/Co interface is opposite to that of the Pt/Co interface (i.e., the magnetic chirality of Pt/Co and Co/Ir interfaces are the same). Additive and large DMI has been observed in Pt/Co/Ir multilayers in which skyrmions are stabilized[8]. Interestingly, however, experiments using magnetic domain walls indicated that the magnetic chirality of the Ir/Co interface is the same as that of the Pt/Co interface[19, 20]. The DMI at the Ir/Co interface thus seems to depend on factors that are yet to be determined. Moreover, in order to electrically control the dynamics of chiral domain walls and skyrmions, it is essential to gain solid understanding on the spin-orbit torque[21-23] (SOT) that drives the chiral magnetic structures.

Here, we study the DMI and SOT in [Pt/Co/Ir]$_N$ multilayers using measurements of the current-induced shift of the anomalous Hall loops[24]. The multilayer stack [Pt/Co/Ir]$_N$ is used since it allows increase in the thermal stability of chiral domain walls and skyrmions via increase in their magnetic volume, which is beneficial for technological applications. We compare the results of [Pt/Co/Ir]$_N$ multilayers with [Pt/Co/Cu]$_N$ multilayers (Cu replacing Ir) to reveal the role the Ir layer plays in defining the DMI and SOT.

**II. Sample preparation and experimental setup**

Films were grown on Si (100) substrates, coated with 100 nm thick silicon oxide, using magnetron sputtering. Multilayer structures are composed of Sub./3 Ta/2 Pt/[0.6 Pt/0.9 Co/$d$ X]$_N$/2 MgO/1 Ta (units in nanometer) with X=Ir or Cu. $N$ represents the number of repeats of the unit structure enclosed by the square brackets. X=Ir, $N$=3 is referred to as film A, X=Cu, $N$=3 is film B and X=Ir, $N$=1 is film C. The thickness (0.1 nm ≤ $d$ ≤ 1.1 nm) of X was varied using a moving shutter during the deposition process. Three reference films were made to characterize the transport properties of Ir and the interface state of Ir/Co: film D: Sub./1.5 Ta/$d$ Ir/1 CoFeB/2 MgO/1 Ta, film E: Sub./1.5 Ta/7 Ir/0.9 Co/2 MgO/1 Ta and film F: Sub./3 Ta/2 Pt/1 Ir/0.9 Co/2 MgO/1 Ta. The thickness of the Ir underlayer in reference film D was varied to determine the spin diffusion length of Ir via spin Hall magnetoresistance (SMR)[25-28] measurements. We use CoFeB as the ferromagnetic layer for the SMR measurements as it has been shown recently that an anomalously large SMR emerges in bilayers with thick Co[29]. Reference films E and F are used to study the effect, if any, of the seed layer of Ir on DMI and SOT at the Ir/Co interface: Ir is grown on highly textured Pt(111)[30] surface for reference film F whereas the seed layer of Ir for films A-C and D is Co and amorphous Ta[31], respectively. (We do not have the information on the structure of the Co layer in films A-C as it is too thin to perform structural characterization.) All films possess sufficiently strong perpendicular magnetic anisotropy so that the magnetic easy axis points along the film normal (i.e. along the $z$-axis). A summary of the film stacking is presented in Table 1.

Optical lithography and Ar ion milling were used to form Hall bars from films A-C, E and F. The width ($w$) and the distance ($L$) between the longitudinal voltage probes are ~10 μm and ~25 μm, respectively. Figure 2(a) shows an optical microscopy image of a typical Hall bar with the definition of the coordinate axis. DC current is applied along the *x*-axis: positive current is defined as current flow to + *x*. External magnetic fields were applied along the *x*, *y*, and *z* directions, referred to as $H_x$, $H_y$, and $H_z$, respectively. DMI, SOT and SMR were evaluated using the patterned Hall bars. For reference film D, Hall bars with *w*~0.4 mm and *L*~1.2 mm were formed using a predefined shadow mask during the deposition process.

Magnetic properties of the films were studied using vibrating sample magnetometer (VSM). The saturation magnetization ($M_s$) and the effective magnetic anisotropy energy ($K_{eff}$) are estimated from the magnetic easy and hard axes hysteresis loops measured. The nominal FM layer thickness is used to calculate $M_s$. The X layer thickness (*d*) dependences of $M_s$ and $K_{eff}$ for films A and B are shown in the Supplementary material[32] (Fig. S1). The results are interpolated to obtain the corresponding value of $M_s$ and $K_{eff}$ for the patterned devices made from films A and B. $M_s$ and $K_{eff}$ of reference films E and F are summarized in Table 2.

### III. Experimental results and discussions

#### A. Spin diffusion length of Ir

We first study the SMR[25-28] of reference film D [Sub./1.5 Ta/*d* Ir/1 CoFeB/2 MgO/1 Ta]

to determine the spin diffusion length of Ir. The longitudinal resistance $R_{xx}$ of the Hall bar was measured while a constant magnitude magnetic field was applied to the sample. The relative angle ($\theta$) between the magnetic field and the film normal was varied, as shown in the inset of Fig. 1(a). The magnetic field was rotated in the $yz$ plane (current flow is along the $x$-axis): under such circumstance the resistance variation against $\theta$ provides information on the SMR[25-27]. The applied magnetic field was large enough (~3 T) to align the magnetic moment of the FM layer (CoFeB) along the magnetic field.

The inset of Fig. 1(b) shows the Ir layer thickness ($d$) variation of the sheet conductance $[L/(w \cdot R_{xx}^z)]$. Except for the thinnest Ir layer film, the sheet conductance scales linearly with $d$. The slope of $L/(w \cdot R_{xx}^z)$ vs. $d$ is proportional to the inverse of the Ir layer resistivity: we estimate the resistivity to be ~19 µΩcm. It is not clear what causes the deviation of the thinnest Ir layer film from the linear fitting: we infer that some degree of intermixing with Ta and/or CoFeB layers may influence the transport properties. The magnetic field angle ($\theta$) dependence of $R_{xx}$ is displayed in Fig. 1(a). Data are fitted with a sinusoidal function to obtain the resistance difference $\Delta R_{xx}^{SMR}$ when the FM layer magnetization points along the $y$-axis ($R_{xx}^y$) and the film normal ($R_{xx}^z$), i.e. $\Delta R_{xx}^{SMR} = R_{xx}^y - R_{xx}^z$. The resistance ratio $\Delta R_{xx}^{SMR}/R_{xx}^z$ is plotted as a function of $d$ in Fig. 1(b). $|\Delta R_{xx}^{SMR}/R_{xx}^z|$ increases with decreasing Ir layer thickness and shows the largest value at an Ir layer thickness of ~2 nm. Based on the theory of SMR[26], the thickness at which $\Delta R_{xx}^{SMR}/R_{xx}^z$ takes a maximum is roughly two times the spin diffusion length of the spin current generating layer. Although the thinnest Ir layer film exhibits a different resistivity from the other films, these results show that the spin diffusion length of Ir is less than ~1 nm. Combination of short spin diffusion length and high conductivity makes Ir a good spin sink.

### B. Current induced shift of the anomalous Hall loop

The anomalous Hall resistance $R_{xy}$ was measured against the out-of-plane field $H_z$ under application of a DC bias current $I_{DC}$ and an in-plane bias field $H_x$. Figure 2(b) shows exemplary $R_{xy}$-$H_z$ loops for a Hall bar made of film C (X=Ir, N=1, d~0.6 nm) and $H_x = 0.2$ T, $I_{DC} = \pm 12$ mA. When positive (negative) current is applied, the center of the hysteresis loop shifts to positive (negative) $H_z$. The shift of the loop center with respect to $H_z = 0$ is defined as $-H_{eff}^z$. $H_{eff}^z$ is plotted as a function of current density (J) in Fig. 2(c). We convert the bias current $I_{DC}$ to $J$ assuming that majority of current flows uniformly in the conducting metallic layers (Pt, Co, Ir). (Taking into account the thickness dependent resistivity of each layer changes estimation of the SOT by at most ~10%.) Since the resistivity of the thin Ta underlayer is nearly an order of magnitude larger than the conducting layers and the MgO/Ta capping layer is insulating (the top Ta layer is oxidized), current flow into these layers is neglected for all structures. As evident, $H_{\text{eff}}^z$ scales linearly with $J$. We thus fit $H_{\text{eff}}^z$ vs. $J$ with a linear function. The slope of the fitted function $H_{\text{eff}}^z/J$ is plotted against $H_x$ in Fig. 2(d).

Following the analyses of Pai et al.[24], the $H_x$ at which $H_{\text{eff}}^z/J$ saturates (Fig. 2(d)) represents the DM exchange field $H_{\text{DM}}$ and the saturation value of $H_{\text{eff}}^z/J$, which we will refer to as $H_{\text{eff}}^z/J|_{\text{sat}}$, is proportional to the spin torque efficiency $\xi_{\text{DL}}$, i.e. $H_{\text{eff}}^z/J|_{\text{sat}} = (\pi/2)(\hbar/2eM_s t_F)\xi_{\text{DL}}$ (see also Ref. [33]). $\hbar$ and $e$ are the reduced Planck constant and the electric charge, respectively, $M_s$ and $t_F$ are the saturation magnetization and the thickness of the FM (Co) layer. Note that the sign of DMI (i.e. the magnetic chirality) cannot be determined from these measurements.

### C. Spin-orbit torque

The $H_x$ dependence of $H_{\text{eff}}^z/J$ for reference films E [Sub./1.5 Ta/7 Ir/0.9 Co/2 MgO/1 Ta] and F [Sub./3 Ta/2 Pt/1 Ir/0.9 Co/2 MgO/1 Ta] are shown in Figs. 3(a) and 3(b), respectively. For both films, the FM layer (Co) is sandwiched between an Ir underlayer and a MgO capping layer. The spin torque efficiency $\xi_{DL}$ is estimated from $H_{eff}^z/J|_{sat}$: the values are listed in Table 2. $\xi_{DL}$ for reference film E is ~0.01. Similar value was reported in Ref. [19]. Since the Ir layer thickness for reference film E is much larger than its spin diffusion length, $\xi_{DL}$ represents the bulk spin Hall angle of Ir (neglecting interfacial effects such as spin memory loss[34]). $\xi_{DL}$ for reference film F is larger than film E due to the larger spin Hall effect of Pt placed below the Ir layer. As discussed below, the Ir layer tends to absorb spin current that diffuses in from neighboring layers, and thus $\xi_{DL}$ of reference film F is likely to be smaller than that of Pt[30, 35]. As the sign of $H_{eff}^z/J|_{sat}$ is the same for films E and F, we consider the spin Hall angles of Ir and Pt possess the same sign.

The spin torque efficiencies $\xi_{DL}$ for films A-C ([0.6 Pt/0.9 Co/$d$ X]$_N$ multilayers) are plotted as a function of X (=Ir and Cu) layer thickness ($d$) in Fig. 4(a). Black squares, red circles and green triangles correspond to $\xi_{DL}$ of film A (X=Ir, $N$=3), film B (X=Cu, $N$=3) and film C (X=Ir, $N$=1), respectively. Comparison of the results from films A and B (X=Ir, Cu, $N$=3) shows that $\xi_{DL}$ for X=Cu multilayers is smaller than that of X=Ir multilayers. To reveal the role of the X layer on $\xi_{DL}$ more precisely, the SMR of films A-C was measured. $\Delta R_{xx}^{SMR}/R_{xx}^z$ is plotted as a function of $d$ in Fig. 4(b). Although the $yz$ plane magnetoresistance may contain contributions from other sources, here we assume the relative magnitude is comparable since the Co layer is sufficiently thin (note the anomalous SMR emerges for thicker Co films[29, 36-39]) and its thickness is the same for both multilayers. For both systems, $|\Delta R_{xx}^{SMR}/R_{xx}^z|$ decreases with increasing $d$. However, $|\Delta R_{xx}^{SMR}/R_{xx}^z|$ is significantly smaller for X=Ir multilayers (film A) compared to that of X=Cu (film B). As the spin accumulation at interfaces is proportional to the spin torque efficiency $\xi_{DL}$, these results are in contrast to the results shown in Fig. 4(a). Note that the single-repeat multilayer ($N$=1, X=Ir, film C) shows similar results with those of the corresponding $N$=3 repeated stacks (film A).

The contradictory results of $\xi_{DL}$ (Fig. 4(a)) and SMR (Fig. 4(b)) can be accounted for qualitatively if we assume the Ir layer acts as a spin sink and hinders spin transmission across the layer. Illustrations of the electron spin transport and the resulting SOT in the unit structure of the multilayers ($N$>1) are depicted in Fig. 5. Figures 5(a) and 5(c) show the spin accumulation $\vec{\sigma}$ at the top and bottom interfaces of the Co layer due to the spin Hall induced spin current generated from the top and bottom Pt layers, respectively. Here we have assumed that Cu and Ir generate negligible spin current (see Table 2 for $\xi_{DL}$ of Ir). Since the bottom Pt/Co interface is the same for both multilayers, the spin torque efficiency depends on the amount of spin accumulation at the top Co interface. For X=Cu (Fig. 5(a)), the spin current from the top Pt layer (on top of Cu) traverses the Cu layer and impinges on the Co layer, resulting in spin accumulation at the Co/Cu interface. Since the spin currents from the top and bottom Pt layers point to opposite directions, the net torque on the magnetic moments will work against each other. $\xi_{DL}$ of X=Ir (Fig. 5(c)) is thus larger than that of X=Cu since the Ir layer absorbs the spin current from the Pt layer due to its short spin diffusion length, which results in reduction of the torque compensation.

The SMR, on the other hand, is proportional to the spin accumulation at the FM/HM interface but the sign of $\vec{\sigma}$ does not influence the overall magnitude: i.e. the SMR scales with the square of the spin Hall angle[26]. The spin accumulation at the top and bottom interfaces therefore contributes to the SMR in a constructive manner, unless the FM layer is too thin to cause cross talk of spin accumulation at the top and bottom interfaces. Assuming negligible cross talk, the larger SMR for the X=Cu multilayer can be accounted for if the degree of spin accumulation at the top interface is larger (see Figs. 5(a-d)). The smaller SMR for the X=Ir multilayers can be attributed to absorption of the spin current at the Ir layer diffusing in from the top Pt layer. This observation is consistent with the results of $N$=1 (X=Ir) multilayers (film C): both $\xi_{DL}$ and the SMR take similar value with those of the $N$=3 multilayers, suggesting that the Pt layer on top of the Ir layer for the $N$=3 multilayer has little influence on the SOT/spin accumulation.

### D. Dzyaloshinskii–Moriya interaction

As schematically shown in Fig. 2(d), $H_{DM}$ is obtained by first fitting $H_{eff}^z/J$ vs. $H_x$ with a linear function in appropriate ranges of $H_x$ around $H_x \sim 0$. We then look for the intersection of the fitted linear line with the saturated value of $H_{eff}^z/J$ and take the $x$-coordinate of the intersection as $H_{DM}$. The DM exchange constant $D$ is calculated from $H_{DM}$ using the following relation $|D| = \mu_0 M_s H_{DM} \Delta$. Here, $A$ is the exchange stiffness constant and $\Delta = \sqrt{A/K_{eff}}$. We assume $A = 15$ pJ/m for all films studied[40, 41]. The obtained $|D|$ for films A, B and C ([0.6 Pt/0.9 Co/$d$ X]$_N$) is plotted as a function of X layer thickness ($d$) in Fig. 6. For films A and B, $|D|$ increases with increasing $d$ until it saturates at a certain $d$. Upon saturation, we find $|D| \sim 1.8$ mJ/m² for X=Cu and $|D| \sim 0.4$ mJ/m² for X=Ir. Interestingly, $|D|$ is significantly larger for X=Cu (similar results have been reported in Ref. [42, 43]). If we assume the DMI at the Co/Cu interface is negligible, the results from X=Cu (film B) suggest the DMI of Pt/Co interface is $|D| \sim 1.8$ mJ/m². The fact that $|D|$ is smaller for X=Ir multilayers (film A) indicates that the DMI of Co/Ir interface has the opposite sign with that of Pt/Co interface, similar to the results reported in Refs. [19, 44]. (In terms of the DMI with the same stacking order, Ir/Co interface and Pt/Co interface possess the same sign.) Assuming that DMI at the top and bottom interfaces of a FM layer are additive, we estimate $|D|\sim 1.4$ or 2.2 mJ/m$^2$ for the Ir/Co interface. (Since the sign of $D$ cannot be determined from these measurements, $|D|$ can be either 1.4 mJ/m$^2$ or 2.2 mJ/m$^2$ to account for the values obtained for the X=Cu and X=Ir multilayers.) The DMI of $N$=1, X=Ir multilayer (film C), shown by the green triangle in Fig. 6, suggest that the number of stacking does not necessarily influence the DMI.

Recently, it was reported that the DMI of the Co/Ir interface may depend on its structure: in particular, the sign of DMI can change between fcc-based and hcp-based Ir structures[18]. These studies suggest that the DMI of the Ir/Co (or Co/Ir) interface can be influenced by the layer underneath it which controls the growth mode. The reference films E and F possess structures in which the underlayer of Ir is different from that of the multilayers (films A-C). The underlayer is Ta and Pt for reference films E and F, respectively, whereas Co is deposited before Ir for the multilayers. The DMI values of the reference films E and F, obtained from the results presented in Fig. 3, are summarized in Table 2. Interestingly, $|D|$ of reference film E (Ir grown on Ta) exhibits similar magnitude ($\sim$1.6 mJ/m$^2$) with that of the Co/Ir interface in the multilayers. We find a smaller $|D|$ for reference film F (Ir on Pt): the origin of the difference in DMI between the reference films E and F is not clear.

We also studied current-induced motion of magnetic domain walls in patterned wires made of films with stacking similar to those of reference films E and F. We find the domain walls move along the current flow in all cases, in agreement with the results reported in Ref. [19]. (Due to strong pinning, it is difficult to move the domain walls along the wire smoothly, which hinders accurate evaluation of the wall velocity.) Since the effective spin Hall angle of Ir has the same sign as that of Pt, these results suggest that the sign of the DMI at the Ir/Co interface is the same as that of the Pt/Co interface. Together with the results presented in Table 2, we conclude that the Ir/Co interface possesses a DMI that has the same sign as that of Pt/Co and the magnitude is similar.

**IV. Summary**

We have studied the spin torque efficiency and the Dzyaloshinskii–Moriya interaction (DMI) at the Co and Ir interface using [Pt/Co/X]$_N$ multilayers (X=Cu and Ir) and Ir/Co, Pt/Ir/Co reference films. The current-induced shift of the anomalous Hall hysteresis loops is used to evaluate the spin torque efficiency and the Dzyaloshinskii-Moriya interaction. We find that Ir possesses a positive and relatively small spin Hall angle of ~0.01 (same sign with that of Pt), and its spin diffusion length is less than ~1 nm. Due to its high electrical conductivity and short spin diffusion length, the Ir layer acts a good spin sink. Such characteristics of Ir can be used to break flows of spin current that will otherwise reduce the spin torque efficiency. The DMI at the interface of Co and Ir is found to be in similar magnitude with that of Co and Pt interface. We find the magnitude of the DM exchange constant at the Ir/Co (and Co/Ir) interface to be ~1.4-2.2 $mJ/m^2$. The sign of the DM exchange constant for Pt/Co and Ir/Co interfaces turns out to be the same, leading to a reduced DMI for the Pt/Co/Ir multilayers. These results show that Ir can be used as an efficient spin absorbing layer as well as a source of DMI.

**Acknowledgements**

The authors thank K. Yawata for technical support. This work was partly supported by JSPS Grant-in-Aid for Scientific Research (16H03853), Specially Promoted Research (15H05702), Casio Science Foundation and the Center of Spintronics Research Network of Japan. Y.-C.L. is supported by JSPS International Fellowship for Research in Japan (Grant No. JP17F17064).

**Figure captions**

**Fig. 1** (a) Longitudinal resistance $R_{xx}$ of reference film D (Sub/1.5 Ta/~2 Ir/1 CoFeB/2 MgO) plotted as a function of the angle $\theta$ between the magnetic field and the film normal (z-axis). The applied magnetic field is 3 T. The inset shows the definition of the coordinate axis. (b) Spin Hall magnetoresistance ($\Delta R_{xx}^{\mathrm{SMR}}/R_{xx}^{z}$) plotted as a function of the Ir layer thickness (d) for reference film D. The inset shows $L/(wR_{xx}^{z})$ vs. d for the same film. The results presented in Fig. 1 were obtained using Hall bars with $L$~1.2 mm and $w$~0.4 mm.

**Fig. 2** (a) Optical micrograph of a representative Hall bar with the definition of the coordinate axis. (b) Anomalous Hall resistance ($R_{xy}$) vs. $H_z$ for two different dc currents ($I_{\mathrm{DC}}=\pm 12$ mA) for film C (X=Ir, N=1, d~0.6 nm). The bias field along x ($H_x$) is fixed to ~0.2 T. Definition of $H_{\mathrm{eff}}^{z}$ is schematically illustrated. (c) $H_{\mathrm{eff}}^{z}$ vs. current density J for the same film with $H_x$ ~±0.2 T. A linear function is fitted to the data to obtain the slope $H_{\mathrm{eff}}^{z}/J$. The fitting results are shown by the solid lines. The error bars show standard deviation of $H_{\mathrm{eff}}^{z}$ from repeated measurements. (d) The slope $H_{\mathrm{eff}}^{z}/J$ plotted as a function of the bias field $H_x$. $H_{\mathrm{DM}}$ and $\xi_{\mathrm{DL}}$ are extracted as schematically drawn (see text for the details).

**Fig. 3** (a,b) $H_{\mathrm{eff}}^{z}/J$ vs $H_x$ for reference films E [Sub/1.5 Ta/~7 Ir/0.8 Co/2 MgO/1 Ta] (a) and F [Sub/3 Ta/2 Pt/1 Ir/0.8 Co/2 MgO/1 Ta] (b).

**Fig. 4** (a) Spin torque efficiency $\xi_{\mathrm{DL}}$ (a) and spin Hall magnetoresistance $\Delta R_{xx}^{\mathrm{SMR}}/R_{xx}^{z}$ (b) plotted as a function of the X layer thickness (d) for films A-C ([0.6 Pt/0.9 Co/d X]$_N$ multilayers). Black squares: film A (X=Ir, N=3), red circles: film B (X=Cu, N=3), and green triangles: film C (X=Ir, N=1). The error bars in (a) represent standard deviation of the data used to obtain $\xi_{\mathrm{DL}}$ from the plot of $H_{\mathrm{eff}}^{z}/J$ vs. $H_x$ with $|H_x| > |H_{\mathrm{DM}}|$ (see e.g. Fig. 2(d)). The error bars for the results shown in (b) are smaller than the symbol size.

**Fig. 5** (a-d) Illustration of spin transport in the [Pt/Co/X]$_N$ multilayers (X=Ir, Cu, N>1). $\vec{m}, \vec{\sigma}$ and $\vec{H}_{\mathrm{SOT}}$ indicate magnetization direction of the Co layer, spin polarization of the conduction electrons drifting from the Pt layers via the spin Hall effect, and the spin-orbit effective field acting on the magnetic moments. $\vec{H}_{\text{SOT}}$ associated with the spin current from the top and bottom Pt layers are illustrated by the red and blue arrows, respectively. (a,b) X=Cu and (c,d) X=Ir. The Ir layer is assumed to absorb spin current that diffuses in from the top Pt layer. Co magnetization points along the film normal, i.e. along $z$, for (a,c) and along the film plane, i.e. along $y$, for (b,d). For the latter, charge current due to the inverse spin Hall effect is depicted by the dotted lines.

**Fig. 6** The DM exchange constant $|D|$ vs. X layer thickness ($d$) for films A-C ([0.6 Pt/0.9 Co/$d$ X]$_N$ multilayers). Black squares: film A (X=Ir, $N$=3), red circles: film B (X=Cu, $N$=3), and green triangles: film C (X=Ir, $N$=1). Broken lines, which are guide to the eye, illustrate values of $|D|$ with large $d$. The error bars represent the range of $H_{\text{DM}}$ when the range of linear fitting to $H^z_{\text{eff}}/J$ vs. $H_x$ is varied.

**Table 1** Summary of the film ID and structure.

| ID | Stack | Note |
|---|---|---|
| A | Sub./3 Ta/2 Pt/[0.6 Pt/0.9 Co/$d$ Ir]$_3$/2 MgO/1 Ta | $d$=0.1~1.1 nm (wedge) |
| B | Sub./3 Ta/2 Pt/[0.6 Pt/0.9 Co/$d$ Cu]$_3$/2 MgO/1 Ta | $d$=0.1~1.1 nm (wedge) |
| C | Sub./3 Ta/2 Pt/[0.6 Pt/0.9 Co/0.6 Ir]$_1$/2 MgO/1 Ta | |
| D | Sub./1.5 Ta/$d$ Ir/1 CoFeB/2 MgO/1 Ta | $d$=1,2,3,4,5,6,7,8 nm |
| E | Sub./1.5 Ta/7 Ir/0.9 Co/2 MgO/1 Ta | |
| F | Sub./3 Ta/2 Pt/1 Ir/0.9 Co/2 MgO/1 Ta | |

**Table 2** Summary of the saturation magnetization $M_s$, the effective magnetic anisotropy energy $K_{eff}$, DMI exchange field $H_{DM}$, the DM exchange constant $|D|$ and the spin torque efficiency $\xi_{DL}$ for [0.6 Pt/0.9 Co/d X]$_N$ multilayers (films A, B, C) and the reference films E and F. For films A and B, we take data from $d\sim0.6$ nm so that the results can be compared to those of film C (note that $|D|$ saturates when $d>\sim0.6$ nm).

| ID | Stack | $M_s$ | $K_{eff}$ | $H_{DM}$ | $|D|$ | $\xi_{DL}$ |
|---|---|---|---|---|---|---|
| unit | | (kA/m) | ($10^5$ J/m$^3$) | (T) | (mJ/m$^2$) | |
| A | [0.6 Pt/0.9 Co/d Ir]$_3$ ($d\sim0.6$ nm) | 1060 | 7.9 | 0.08 | 0.4 | 0.08 |
| B | [0.6 Pt/0.9 Co/d Cu]$_3$ ($d\sim0.6$ nm) | 1490 | 3.2 | 0.18 | 1.8 | 0.07 |
| C | [0.6 Pt/0.9 Co/0.6 Ir]$_1$ | 1060[a] | 4.7[b] | 0.11 | 0.7 | 0.07 |
| E | 7 Ir/0.9 Co | 910 | 2.3 | 0.22 | 1.6 | 0.01 |
| F | 2 Pt/1 Ir/0.9 Co | 950 | 2.9 | 0.15 | 1.1 | 0.03 |

(a) $M_s$ is assumed to be the same with that of film A (X=Ir, N=3) with $d\sim0.6$ nm.
(b) $K_{eff}$ obtained from $K_{eff} = \mu_0 H_K M_s/2$ where the magnetic anisotropy field ($H_K$) was measured from the in-plane magnetic field dependence of the anomalous Hall resistance.

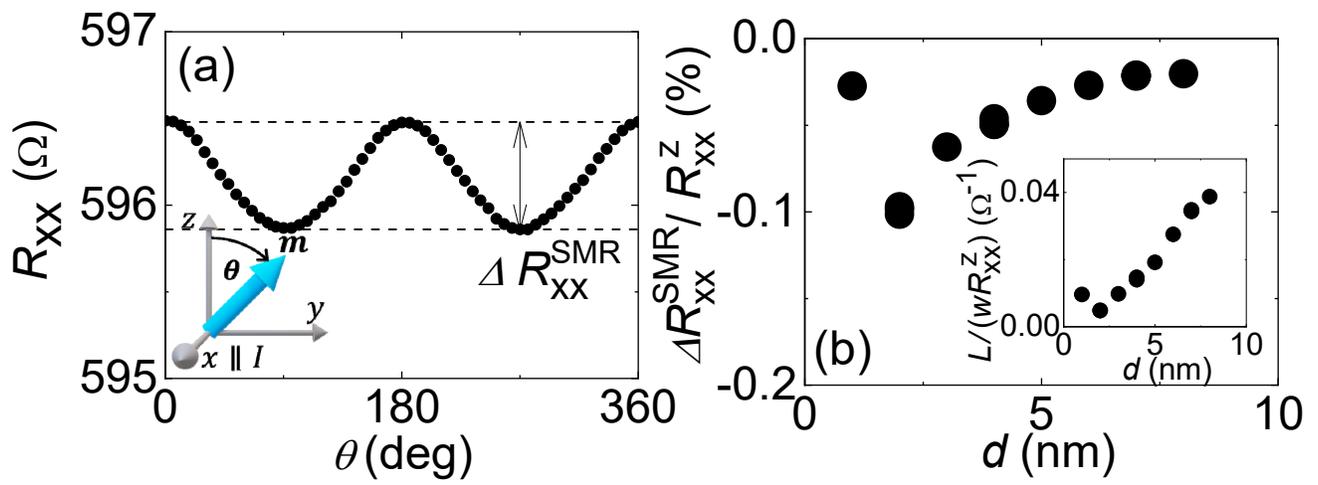

**Fig. 1**

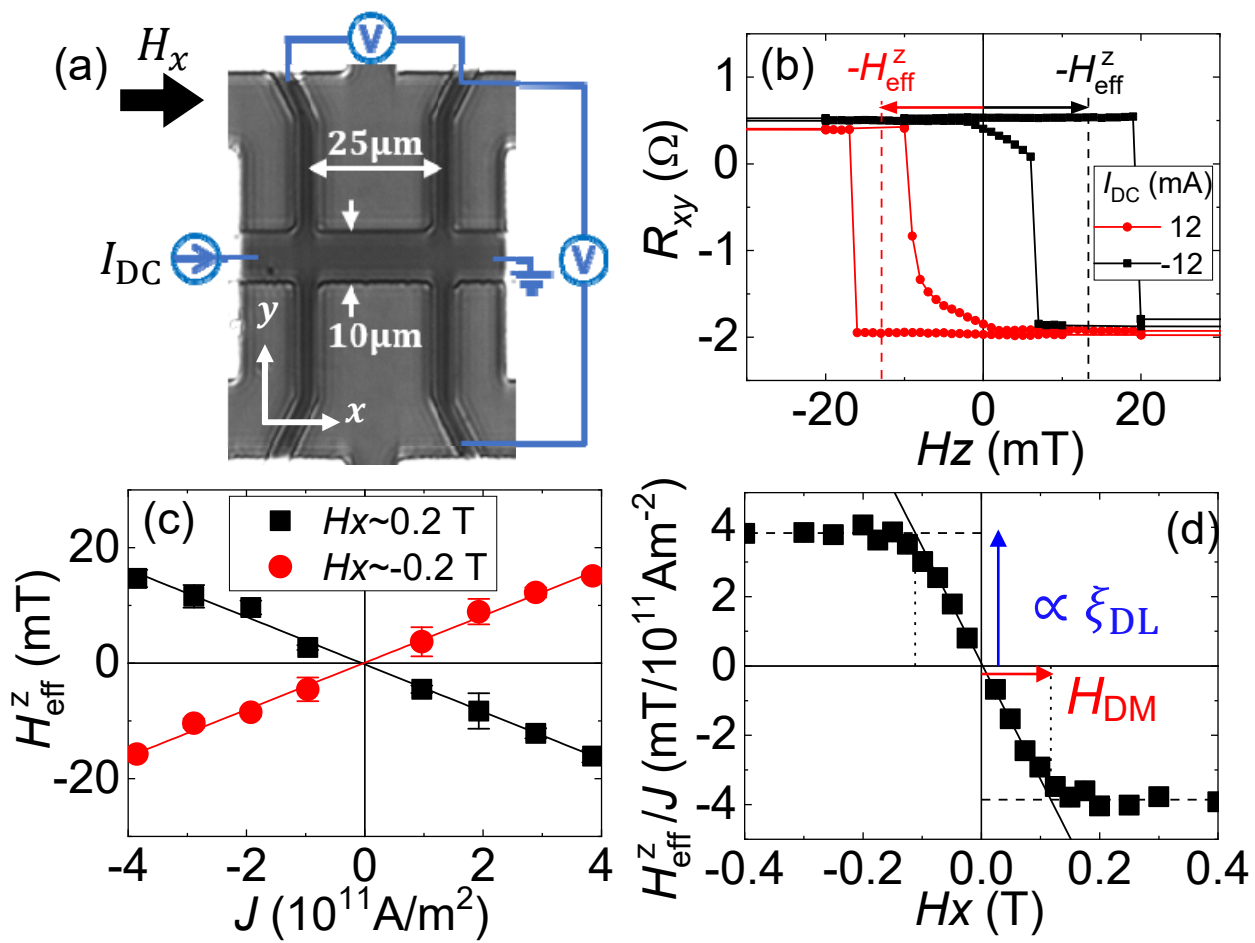

**Fig. 2**

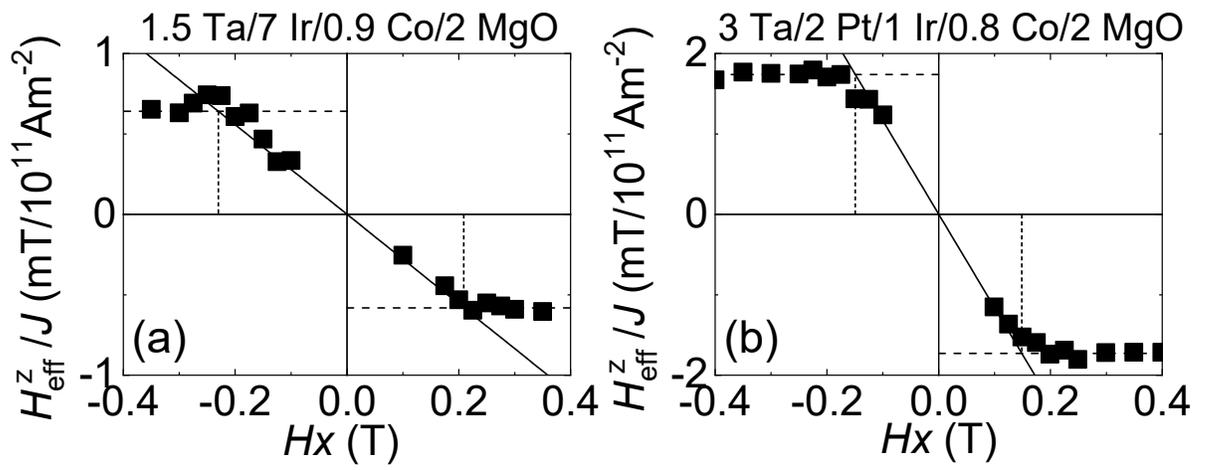

**Fig. 3**

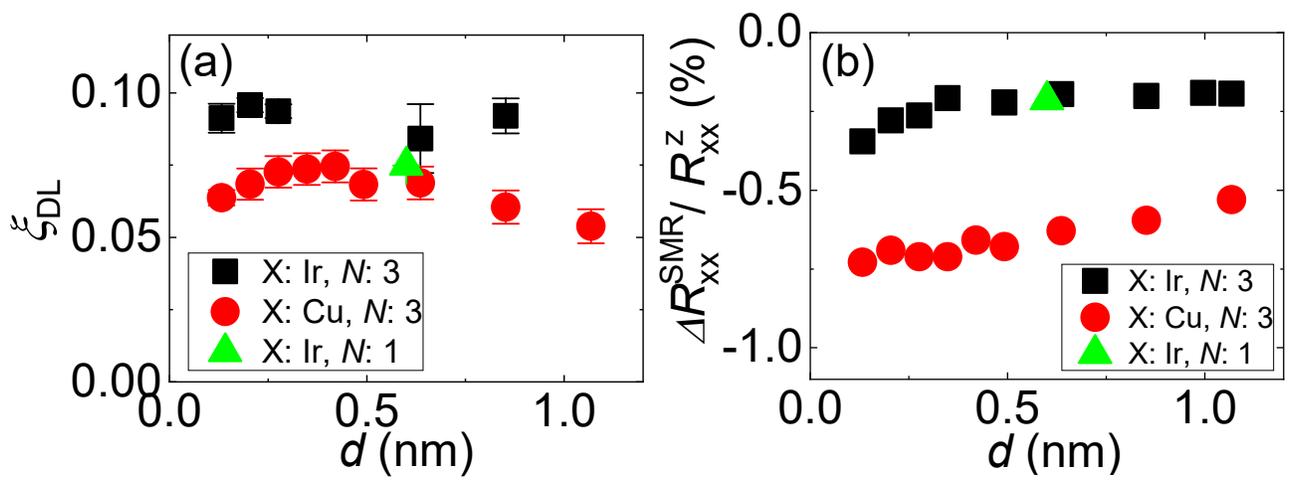

**Fig. 4**

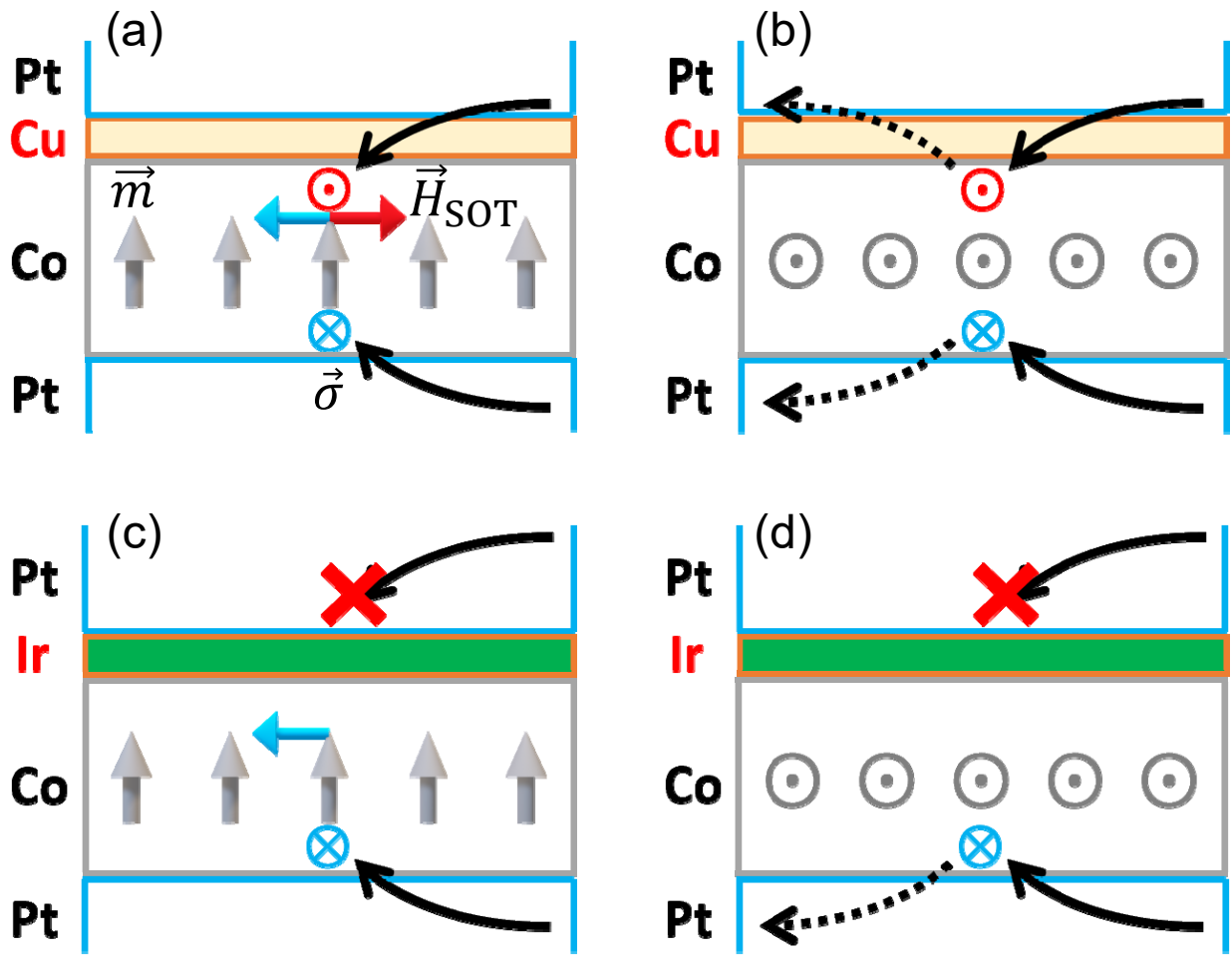

**Fig. 5**

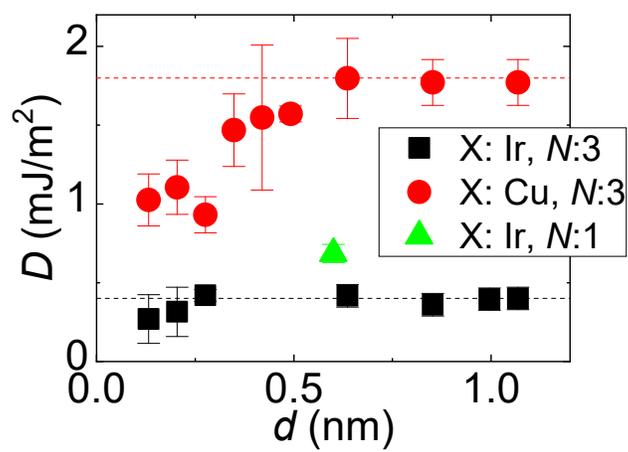

**Fig. 6**

# Supplementary material for

# Dzyaloshinskii–Moriya interaction and spin-orbit torque at the Ir/Co interface

Yuto Ishikuro[1], Masashi Kawaguchi[1], Naoaki Kato[1], Yong-Chang Lau[1,2], and Masamitsu Hayashi[1,2*]

[1]Department of Physics, The University of Tokyo, Bunkyo, Tokyo 113-0033, Japan
[2]National Institute for Materials Science, Tsukuba 305-0047, Japan

## 1. Magnetic properties of the multilayer films

The saturation magnetization ($M_s$) and the effective magnetic anisotropy energy ($K_{eff}$) for films A and B are plotted against the X layer thickness ($d$) in Fig. S1. $K_{eff}$ is obtained by calculating the areal difference of the out-of-plane and in-plane magnetization hysteresis loops.

**Figure captions**

**Fig. S1** The X layer thickness ($d$) dependence of the saturation magnetization ($M_s$) and the effective magnetic anisotropy energy ($K_{eff}$) of films A and B.

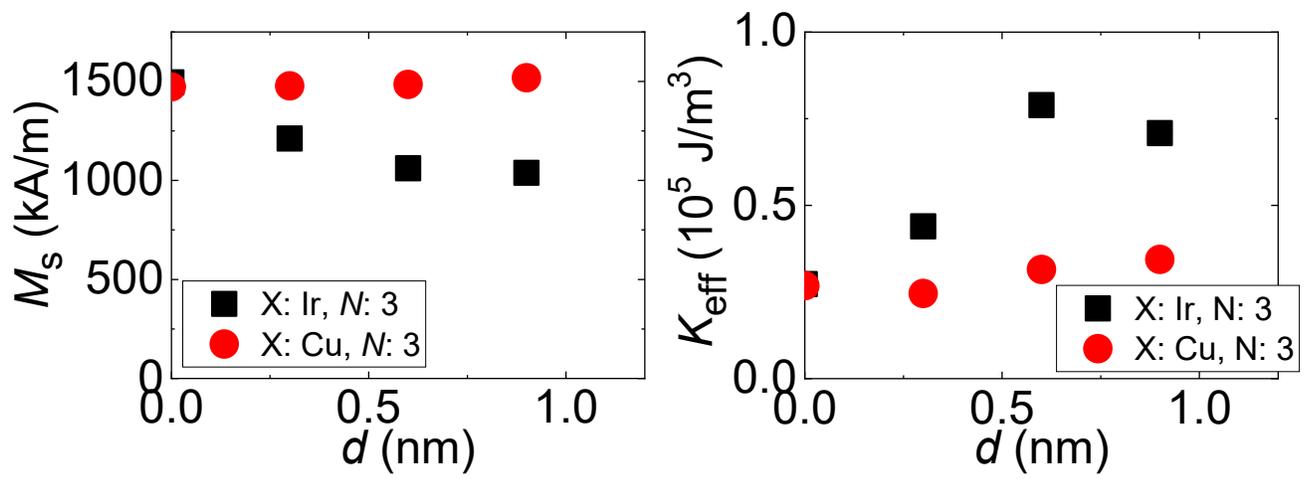

**Fig. S1**